\documentclass{article}

\usepackage{PRIMEarxiv}

\usepackage[utf8]{inputenc} 
\usepackage[T1]{fontenc}    
\usepackage{hyperref}       
\usepackage{url}            
\usepackage{booktabs}       
\usepackage{amsfonts}       
\usepackage{nicefrac}       
\usepackage{microtype}      
\usepackage{lipsum}
\usepackage{fancyhdr}       
\usepackage{graphicx}       
\usepackage{siunitx}
\usepackage{amsmath}
\graphicspath{{media/}}     

\title{The relationship between axial resolution and signal-to-noise ratio in optical coherence tomography
}

\author{
  Danielle J. Harper \\
  Wellman Center for Photomedicine, Massachusetts General Hospital, 40 Blossom Street, Boston, MA 02114, USA \\
  Harvard Medical School, 25 Shattuck Street, Boston, MA 02115, USA \\
  \texttt{djharper@mgh.harvard.edu} \\
   \And
  Benjamin J. Vakoc \\
  Wellman Center for Photomedicine, Massachusetts General Hospital, 40 Blossom Street, Boston, MA 02114, USA \\
  Harvard Medical School, 25 Shattuck Street, Boston, MA 02115, USA\\Harvard-MIT Division of Health Sciences and Technology, 77 Massachusetts Avenue, Cambridge, MA 02139, USA \\
  \texttt{bvakoc@mgh.harvard.edu} \\
}

\begin{document}
\maketitle

\begin{abstract}
In optical coherence tomography (OCT), axial resolution and signal-to-noise ratio (SNR) are typically viewed as uncoupled parameters. We show that this is only true for mirror-like surfaces, and that in diffuse scattering samples such as tissue there is an inherent coupling between axial resolution and measurement SNR. We explain the origin of this coupling and demonstrate that it can be used to achieve increased imaging penetration depth at the expense of resolution. Finally, we argue that this coupling should be considered during OCT system design processes that seek to balance competing needs of resolution, sensitivity, and system/source complexity.
\end{abstract}

\vspace{20pt}
The inverse relationship between optical bandwidth and imaging axial resolution is a defining property of OCT. In this work, we ask whether the optical bandwidth also affects the imaging signal-to-noise ratio (SNR). In asking this question, we put aside experimental complications that scale with source bandwidth, such as those related to source design and interferometer polarization-mode dispersion, and we focus on the fundamental relationship between optical bandwidth and SNR. Stated in more practical terms, this work asks whether there are scenarios in which one should purposely limit axial resolution performance to enhance imaging SNR. 

We conclude that optical bandwidth/axial resolution does influence SNR, but that this is conditional on the nature of the sample. For specular reflections such as those from mirrors, there is no dependence of SNR on optical bandwidth/axial resolution, whereas for diffuse scattering samples, such as biological tissues, there is a fundamental inverse relationship between optical bandwidth and measurement SNR. Given that OCT is predominantly used to image diffuse scattering samples, we argue that this relationship should at minimum be appreciated, and that in some applications it should be considered during the design of the OCT system. We also show that it is because of this relationship that split-spectrum methods in OCT (e.g, spectral-binning polarization-sensitive OCT) are as effective as they are. 

First we present a physical argument supporting the assertion that, in diffuse scattering samples, measurement SNR is inversely proportional to the optical bandwidth used to make the measurement. Consider two swept-source OCT systems (the argument applies also to spectral-domain architectures) with equivalent optical power and A-line durations, but with one sweeping over 100 \si{\nano\meter} optical bandwidth and the other over only 10 \si{\nano\meter}. For simplicity, we assume that each system uses a digitizer to capture its fringe with a fixed number of samplings during the A-line. The detectors and detector bandwidths are identical. For each, the noise performance is determined by optical noise in the reference arm and is therefore equivalent, i.e., the optical bandwidth over which the laser traces has no impact on the measured noise. The signal power within each measurement voxel, or equivalently each bin of the eventual discrete Fourier transform of the digitized fringe, scales inversely with optical bandwidth. This is because lower optical bandwidths capture a larger number of scattered photons from the diffusely scattering sample. The 100 \si{\nano\meter} system, for example, captures the backscattered light within an approximately 7 µm depth range, while the 10 \si{\nano\meter} system captures the backscattered light within a 70 µm depth range. There are 10$\times$ more scatterers within the 70 µm range than the 7 µm voxel (assuming constant reflectivity), leading to 10$\times$ higher sample arm power in each voxel, and a 10-fold higher SNR (10 dB). Stated concisely, this argument predicts that, for a diffuse scattering sample, the relationship between the measurement SNR of a voxel and measurement optical bandwidth ($\Delta\nu$) is given as $\text{SNR}_{\text{diffuse}} \propto (\Delta\nu)^{-1}$.

The same reasoning can be applied to a mirror, specular reflection, or other sub-axial resolution boundary to predict that SNR is independent of optical bandwidth, i.e.,  $\text{SNR}_{\text{specular}} \propto (\Delta\nu)^0$. For these samples, all scatterers are co-localized at a single depth location, and so extending the measurement voxel depth extent does not admit additional scatterers into the voxel, and therefore does not increase signal power. 

This relationship can be seen in the conventional mathematical description of OCT if we are careful to describe specular and diffusely scattering samples accurately. We take as a starting point the cross-correlation term of an OCT A-line, as in Ref.~\cite{izatt2008theory}. For a mirror sample, the OCT signal is emitted from a single reflector with field reflectivity $r_{s}$ located at an axial position $z_{s}$, and the cross-correlation term, $i_{D\star}$, is confined by the delta function, $\delta$, such that

\begin{equation}
\label{eq:1}
i_{D\star}(z)=\frac{\rho}{4}\left[\gamma(z) \otimes r_{r} r_{s}\delta\left(z \pm 2 z_{s}\right)\right]
\end{equation}

\noindent where $\rho$ is the responsivity of the detector, $r_{r}$ is the field reflectivity of the reference arm, and we have assumed the reference arm is located at $z=0$. The parameter $\gamma(z)$ is the coherence function of the source and is normalized to $\gamma(0) = 1$. The width of $\gamma(z)$ describes the axial resolution of the system. As expected, the measured signal at the mirror position, $i_{D\star}(2z_{z})$, does not depend on the width of $\gamma(z)$ and therefore does not depend on the optical bandwidth of the source.  

For a diffusely scattering sample, the signal is a contribution from all of the scatterers across depth, each convolved with the coherence function. If we consider $N$ scatterers located across a depth range that is larger than the width of the coherence function, the cross-correlation term becomes
\begin{equation}
\label{eq:2}
i_{D\star}(z)=\frac{\rho}{4}\left[\gamma(z) \otimes \sum_{n=1}^{N} r_{r} r_{s n}\delta\left(z \pm 2z_{s n}\right)\right].
\end{equation}

\noindent For simplicity, we will consider the scatterer reflectivities $r_{sn}$ as real-valued, both positive and negative. Consistent with the diffuse scattering regime, we assume that there are numerous scatterers within the width of the coherence function. Assuming we are sufficiently deep within the sample that there are no surface effects to consider, the signal $i_{D\star}(z)$ is the sum of multiple scatterers. The number of scatterers that contribute to this sum scales with the width of the coherence function; a wider coherence function admits more scatterers. The ensemble average of this signal, $\langle i_{D\star}(z)\rangle$, then scales as the square root of the number of scatterers (random walk), which implies it scales with the square root of the width of $\gamma(z)$. The corresponding measurement intensity, $\langle|i_{D\star}(z)|^2\rangle$, is then directly proportional to the width of $\gamma(z)$ and so inversely proportional to $(\Delta\nu)$. Note that this relationship is exact only for cases wherein the source spectral shape (e.g., Gaussian) remains unchanged. 

It is interesting to note that, if we apply a similar mathematical reasoning to the autocorrelation terms (which are not included in Eqs.~\ref{eq:1} and~\ref{eq:2}), the SNR of the autocorrelation signal intensities is predicted to be inversely proportional to $(\Delta\nu)^{2}$. 

\begin{figure}[htbp!]
\centering\includegraphics[width=6cm]{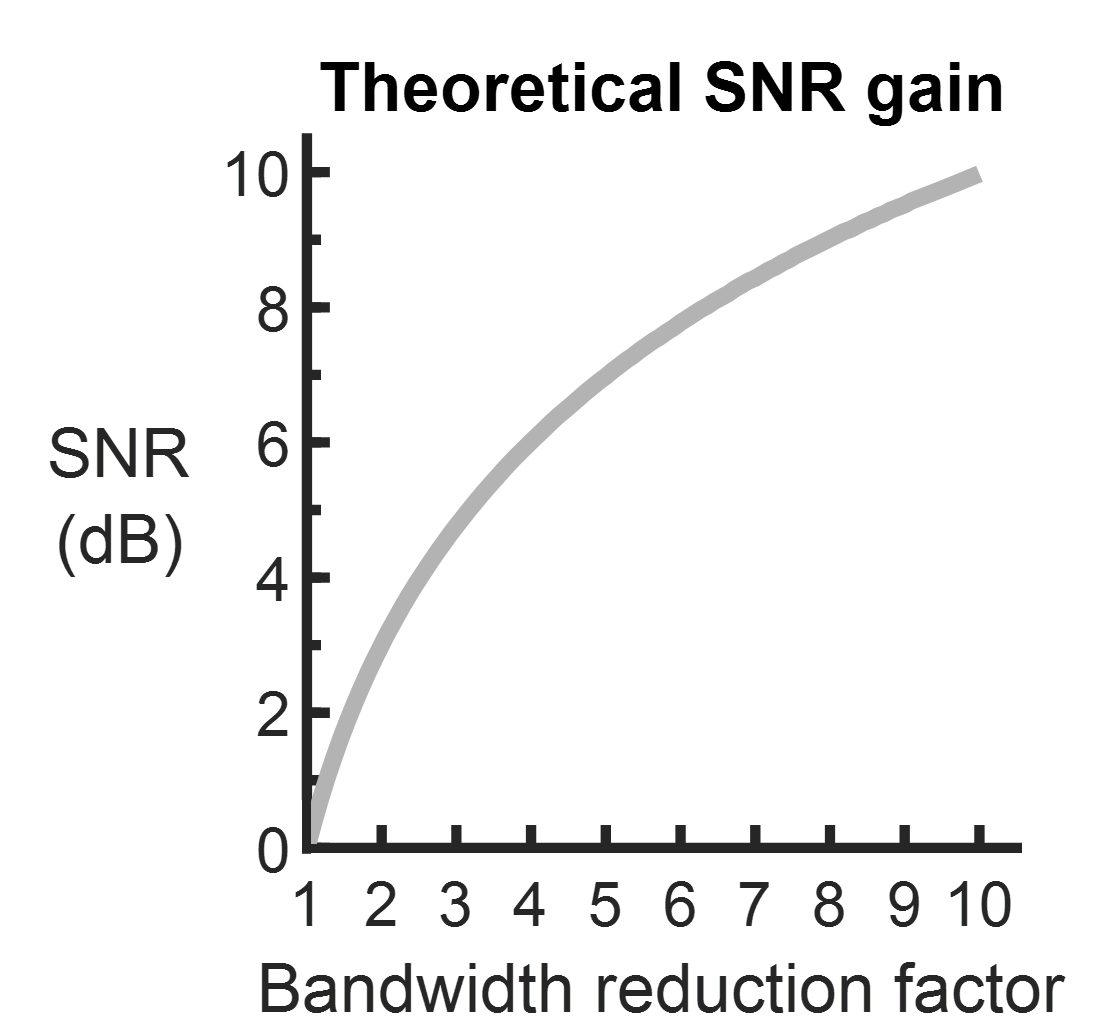}
\caption{Theoretical prediction of diffuse scatterer SNR gain due to a reduction in optical bandwidth (logarithmic scale).}
\label{fig1}
\end{figure}

The $\text{SNR}_{\text{diffuse}} \propto (\Delta\nu)^{-1}$ relationship is shown in logarithmic  scale in Fig. \ref{fig1}. Additionally, it is instructive to explicitly include the A-line duration, $\tau$, into these SNR relationships. It is well known that the measurement SNR, regardless of sample type, scales linearly with $\tau$, yielding  $\text{SNR}_{\text{diffuse}} \propto (\tau)(\Delta\nu)^{-1}$ and $\text{SNR}_{\text{specular}} \propto (\tau)(\Delta\nu)^0$.

\begin{figure}[ht!]
\centering\includegraphics[width=9.0cm]{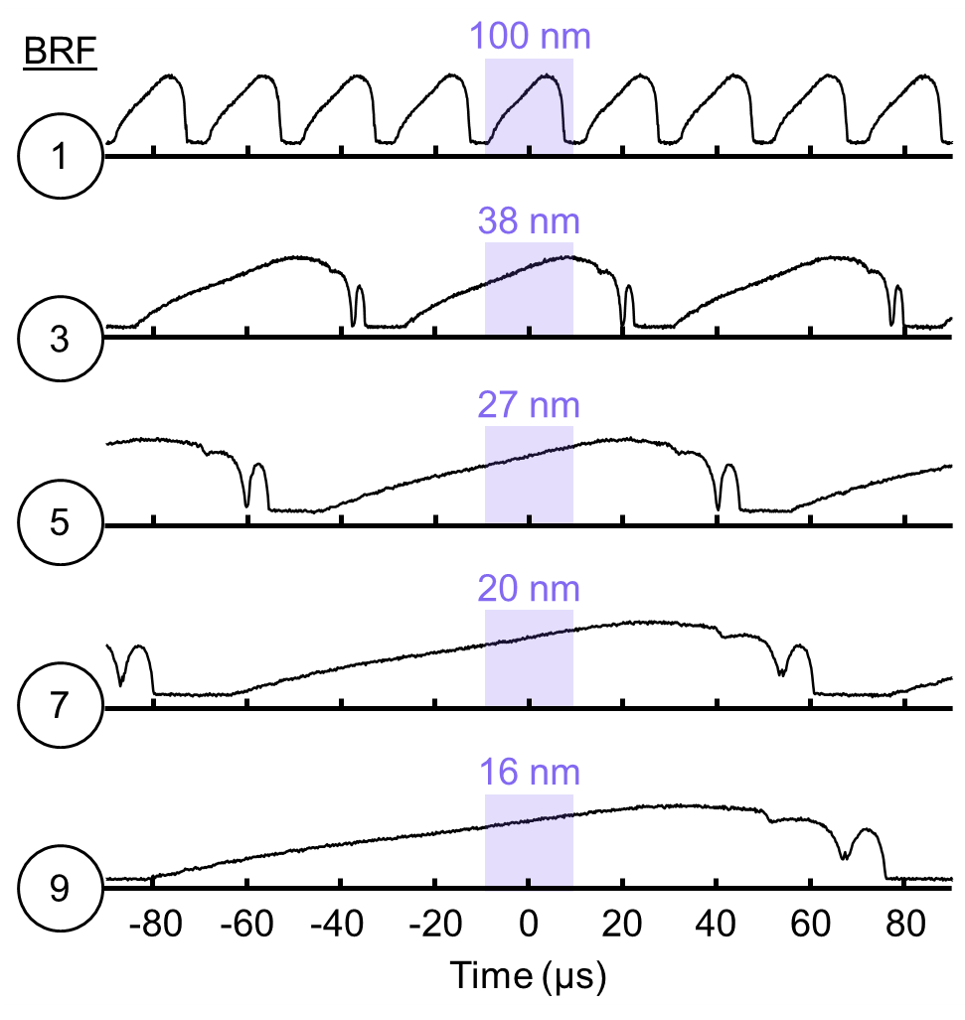}
\caption{Sampled spectra measured by an oscilloscope. The numbers in the left column represent the bandwidth reduction factor (BRF) controlled by the polygon mirror angular velocity and the shaded regions indicate the optical bandwidths associated with a 20 \si{\micro\s} acquisition. Trigger signals were shifted to maintain the same center wavelength.}
\label{fig2}
\end{figure}

\begin{figure*}[htb]
\centering\includegraphics[width=17cm]{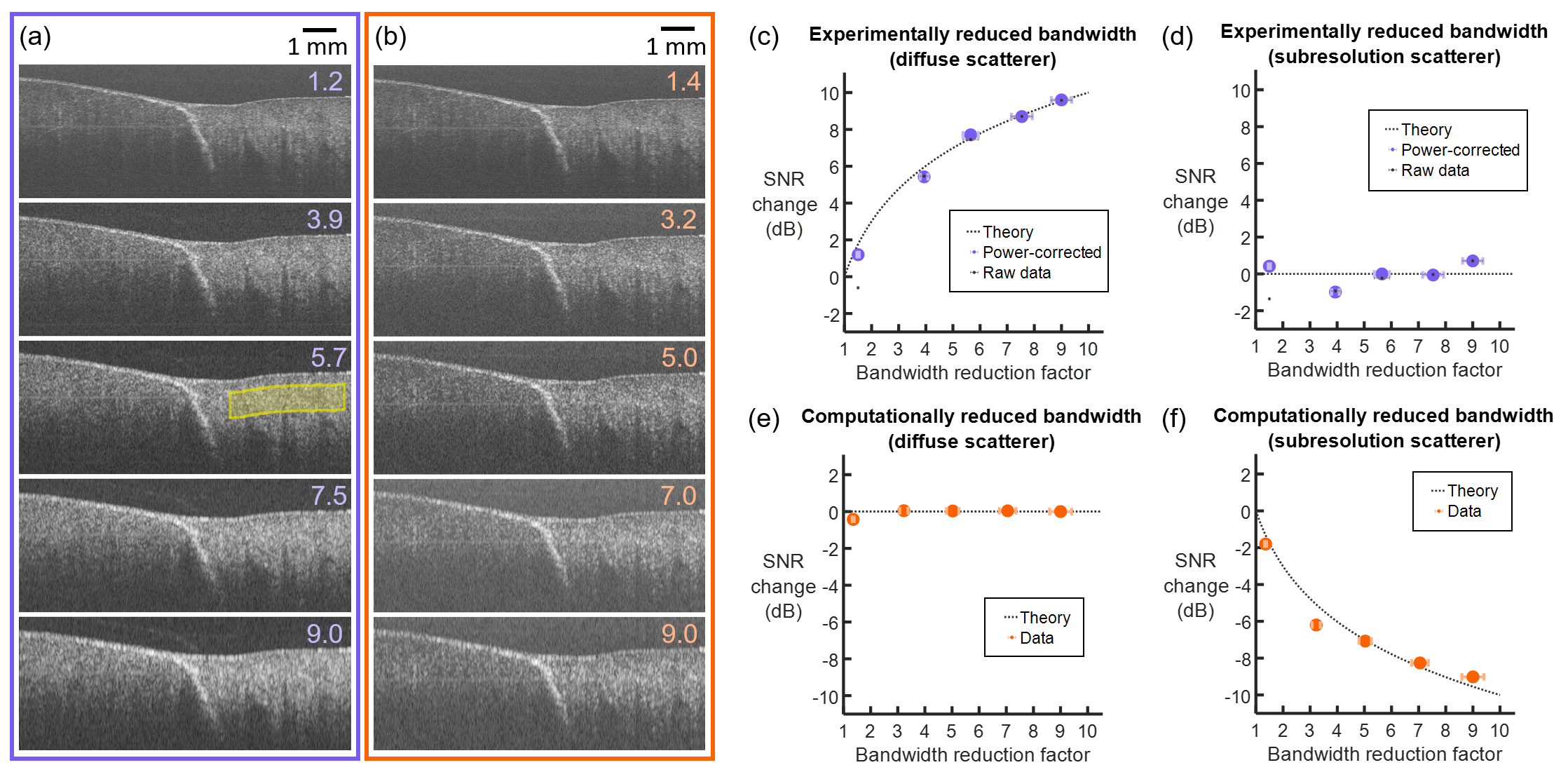}
\caption{Qualitative chicken breast imaging results. a) Image series generated by experimental optical bandwidth reduction. b) Image series generated by computational optical bandwidth reduction. c) SNR as a function of experimental bandwidth reduction factor for the area indicated with the yellow box in (a). d) SNR as a function of experimental bandwidth reduction factor for a mirror signal. e) SNR as a function of computational bandwidth reduction factor for the area indicated with the yellow box in (a). f) SNR as a function of computational bandwidth reduction factor for a mirror signal.}
\label{fig3}
\end{figure*}

To validate these relationships, we performed a series of imaging experiments using a swept source OCT system that allowed tuning of the optical bandwidth while holding the A-line duration constant and the average power approximately constant. The system was based on a polygon scanning mirror and was centered at 1.3 \si{\micro\m} \cite{vakoc2005phase,vakoc2009three}. The angular velocity of the polygon scanning mirror can be tuned over a large range, which modulates the A-line rate of the source correspondingly as shown in Fig.~\ref{fig2}. We reduced the rotation frequency of the polygon mirror by factors of 1 (full resolution, 50 kHz sweep rate), 3 (17 kHz sweep rate), 5, (10 kHz sweep rate), 7 (7.1 kHz sweep rate,) and 9 (5.6 kHz sweep rate). We maintained the same A-line acquisition clock and A-line sample count (2048 unique sampling points over $\tau=$ 20 \si{\micro\s} per A-line) for each acquisition. A small variation in optical power was induced by the spectral envelope of the source, primarily between the full bandwidth setting and all of the reduced bandwidth settings. A measurement of the spectral shape from the oscilloscope allowed us to correct for this variation. Laser chirp and system dispersion correction curves were calculated for each bandwidth  configuration. A Hann window was applied prior to Fourier transformation to remove side lobes \cite{tripathi2002spectral}. A long working distance scan lens with an effective focal length of 54 mm (LSM54-1310, Thorlabs) was used as the microscope objective in the system. This minimized variations in measurement SNR due to beam focusing. 

Images of chicken breast were acquired and the SNR performance was analyzed at subsurface regions representative of diffuse scattering. For comparison, images of a mirror surface (representative of a specular reflection) were also acquired and analyzed. The SNR was evaluated for a constant A-line duration and varying optical bandwidths by physically changing the polygon speed as described above (Fig. \ref{fig3}(a)), and also for a fixed polygon rotation frequency (50 kHz A-line rate) but with the fringe duration ($\tau$) reduced in post-processing to achieve the same optical bandwidth (e.g., spectral windowing, Fig. \ref{fig3}(b)).

\begin{figure}[ht!]
\centering\includegraphics[width=9.0cm]{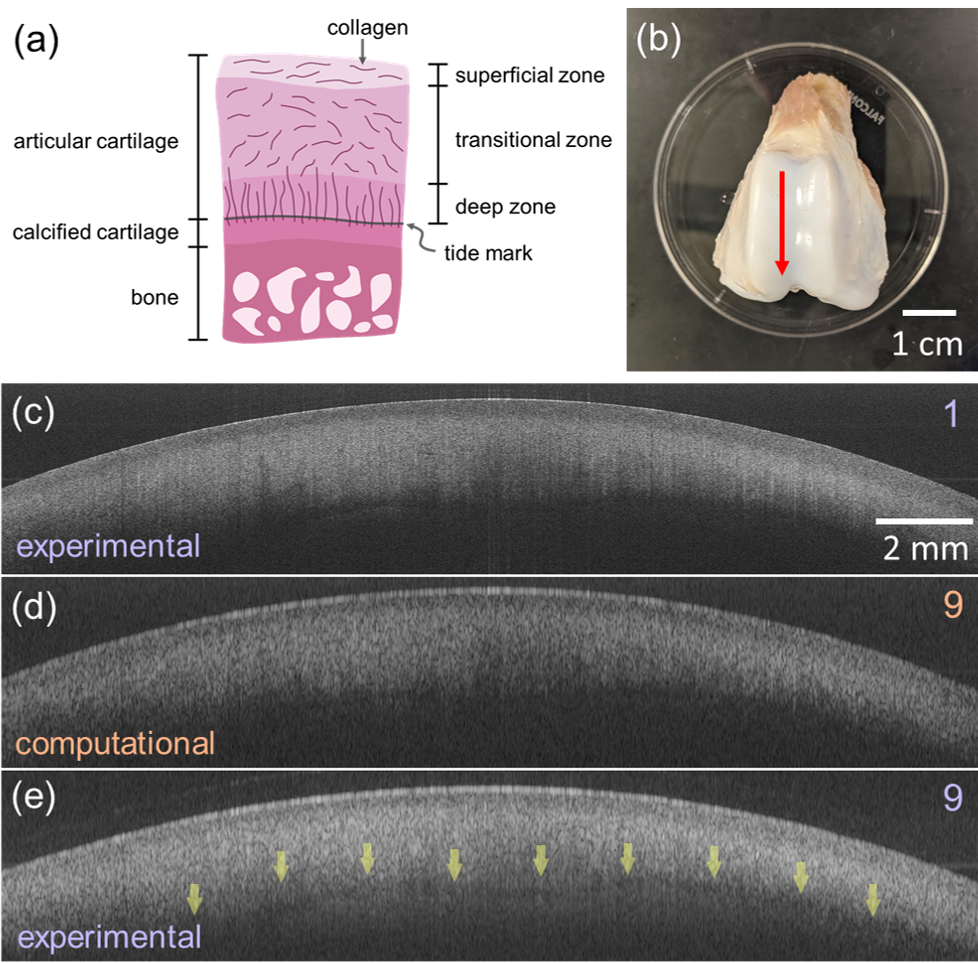}
\caption{Results of porcine knee cartilage imaging. a) Diagram of the structure of the porcine knee. b) Color photograph of the imaged region. Red line indicates the location of the images in (c-e), and the arrow indicates the direction of scanning [left to right on (c-e)]. c) Full resolution OCT image. d-e) Computationally (d) and experimentally (e) generated 9$\times$ bandwidth reduction. Yellow arrows indicate the tide mark between the articular cartilage and the subchondral bone, which is only visible by acquiring with reduced bandwidth.}
\label{fig4}
\end{figure}

The results of this experiment confirm the scaling relationships described above. Figure.~\ref{fig3}(c) demonstrates that SNR increases in diffuse scattering regions in proportion to the inverse of the optical bandwidth. Figure~\ref{fig3}(d) demonstrates that the SNR gain seen in Fig.~\ref{fig3}(c) derives from the diffuse scattering feature of the sample. It is important to note that the mirror signals that are commonly used to characterize OCT system resolution and sensitivity would follow the scaling of Fig.~\ref{fig3}(d) and not demonstrate a dependence on optical bandwidth. The axial resolution of an OCT system and the SNR of the images it produces, therefore, cannot be decoupled from one another, and the measured sensitivity must be used carefully as a predictor of imaging SNR. Figure~\ref{fig3}(e) shows that the SNR in diffuse regions remains largely unchanged when the optical bandwidth is computationally reduced. This is interesting because it is representative of spectral windowing processing approaches wherein a narrow portion of the acquired fringe is processed, causing the optical bandwidth and the effective A-line duration to decrease. The two resultant effects (more photons within the coherence gate due to resolution degradation and fewer detected photons due to reduction of measurement duration) cancel to yield constant SNR. While it has never been explicitly stated, this is in fact why these approaches can be used in split-spectrum angiography \cite{jia2012split}, spectral-binning polarization-sensitive OCT \cite{villiger2013spectral}, frequency compounding for speckle reduction \cite{schmitt1999speckle}, and spectroscopic \cite{morgner2000spectroscopic} or hyperspectral \cite{harper2019hyperspectral} OCT without SNR penalty (within the diffuse scattering regime). Figure~\ref{fig3}(f) shows that for a specular reflector, however, a SNR penalty will be observed when these windowing methods are applied. Herein lies the SNR/axial resolution trade-off, and next we demonstrate how the reduction of optical bandwidth can be used to achieve deeper imaging penetration at the expense of axial resolution in porcine cartilage (Fig. \ref{fig4}).

Porcine knee cartilage is a relatively thick ($\sim$2 mm) and fairly homogeneous tissue with a cartilage-bone interface that is often at the edge of detectability by OCT (Fig.~\ref{fig4}(a)) \cite{jelly2019optical}. In a scenario where the articular cartilage thickness is the critical parameter, our results above would argue that it may be advantageous to image with a dramatically reduced optical bandwidth to prioritize SNR performance. We imaged a knee sample (color photograph in Fig.~\ref{fig4}(b)) using the same protocol as described above. Figure \ref{fig4}(c) shows a cross-sectional image for a full resolution acquisition, while Figs. \ref{fig4}(d) and \ref{fig4}(e) demonstrate 1/9 optical bandwidths from computationally- and experimentally-windowed data, respectively. It is clear that experimental acquisition with 1/9 bandwidth provides the best visibility of the tide mark boundary. Additional applications in which this may be useful include tumor boundary demarcation \cite{vakoc2009three} or industrial tablet coating thickness measurements \cite{koller2011non}.

At the other end of the SNR-resolution optimization continuum, if the goal of a study was to investigate the microstructure of a material containing many boundary-like features, then the axial resolution is of utmost importance. In that case, the SNR of those boundary features will not be compromised, assuming they are specular, but a region of more homogenous scattering would be. This phenomenon may also play a role in contrast improvement between these boundary-type signals and distributed scatterer signals when measured with high resolution OCT. In a recent manuscript by Pi. et al., a high resolution (1.2 \si{\micro\meter}) visible light OCT system visualized highly scattering cell bodies located within the traditionally less-scattering nuclear layers of the retina \cite{pi2020imaging}. Our results suggest that since these cell bodies appear approximately on the scale of the axial resolution of the system, their SNR is not affected by the higher resolution. In contrast, the background of the nuclear layer does see an SNR reduction, and so the "cell-body-to-nuclear-layer" intensity ratio increases when imaged by the higher resolution system. Further study would be required to verify this effect as the source of heightened contrast.   

With recent advances of laser technology allowing for partially coherent light across a large spectral bandwidth, there has been a drive within the OCT community towards increasingly high axial resolution. Until now, the sensitivity measurements performed on high resolution systems have been compared to those on lower resolution systems on a like-for-like basis. Many systems which report axial resolutions on the order of 1 \si{\micro\meter} are based on supercontinuum light sources which have been accepted to be noisy \cite{jensen2019noise}. Even though broad bandwidth OCT systems based on supercontinuum light sources are approaching shot-noise limited sensitivity \cite{ds2021shot}, this work outlines that they will continue to produce images with lower image SNR than their more limited optical bandwidth counterparts (assuming equivalent lateral resolution).

There are also some physical limitations which will limit the achievable SNR enhancement. As the axial resolution approaches the size of the sample structure under investigation, the distributed scatterer approximation no longer holds, and the entire structure then falls into the surface boundary regime wherein SNR is independent of optical bandwidth/axial resolution. There is also a limit to the SNR improvement that can be achieved by degrading axial resolution through bandwidth reduction. This limit can result from the depth-dependent attenuation of the signal, which diminishes the signals from the lower portion of the voxel relative to the upper portion, or from the confocal gate. The later is of course most relevant to optical coherence microscopy systems. Finally, it should be remembered that this SNR improvement is a result of physically reducing the optical bandwidth of the source. The inaccurate compensation of dispersion and/or wavenumber chirp would degrade the axial resolution, but not provide enhanced SNR. After a careful consideration of these limitations and an assessment of study goals, is likely that some OCT systems are better operated with deliberately reduced axial resolution.  

\section*{Funding}
National Institute of Health (P41 EB015903). U.S. Department of Defense through the Military Medical Photonics Program (FA9550-20-1-0063). 

\section*{Acknowledgments}
The authors would like to thank Mohsen Erfanzadeh and Robert Redmond for assistance with preparation of the porcine knee sample. Thanks are also extended to Yong-Chul Yoon, Yongjoo Kim, Tae Shik Kim, Ahhyun Stephanie Nam, Norman Lippok, Hyun-Sang Park, N\'{e}stor Uribe-Patarroyo and Martin Villiger for thought-provoking discussions and feedback throughout. Portions of this work were presented at the European Conference on Biomedical Optics (ECBO) in 2021, paper EW3C–7 \cite{harper2021image}.

\section*{Disclosures}
The authors declare no conflicts of interest. 

\section*{Data Availability}
Data underlying the results presented in this paper may be obtained from the authors upon reasonable request.

\bibliographystyle{unsrt}  
\bibliography{sample}

\end{document}